# Achieving Adaptation for Adaptive Systems via Runtime Verification: A Model-Driven Approach

Zhuoqun Yang[1], Zhi Jin[2], Zhi Li[3]

[1]Institute of Mathematics, Academy of Mathematics and Systems Science, Chinese Academy of Sciences,
Haidian Dstr., Beijing 100190, P. R. China

[2]Key Laboratory of High Confidence Software Technologies (MoE), Peking University,
Haidian Dstr., Beijing 100871, P. R. China

[3]Software Engineering Dept., College of Computer Science and Information Technology, Guangxi Normal University,
Guilin, Guangxi 541004, P. R. China

**Abstract**
Self-adaptive systems (SASs) are capable of adjusting its behavior in response to meaningful changes in the operational context and itself. The adaptation needs to be performed automatically through self-managed reactions and decision-making processes at runtime. To support this kind of automatic behavior, SASs must be endowed by a rich runtime support that can detect requirements violations and reason about adaptation decisions. Requirements Engineering for SASs primarily aims to model adaptation logic and mechanisms. Requirements models will guide the design decisions and runtime behaviors of system-to-be. This paper proposes a model-driven approach for achieving adaptation against non-functional requirements (NFRs), i.e. reliability and performances. The approach begins with the models in RE stage and provides runtime support for self-adaptation. We capture adaptation mechanisms as graphical elements in the goal model. By assigning reliability and performance attributes to related system tasks, we derive the tagged sequential diagram for specifying the reliability and performances of system behaviors. To formalize system behavior, we transform the requirements model to the corresponding behavior model, expressed by Label Transition Systems (LTS). To analyze the reliability requirements and performance requirements, we merged the sequential diagram and LTS to a variable Discrete-Time Markov Chains (DTMC) and a variable Continuous-Time Markov Chains (CTMC) respectively. Adaptation candidates are characterized by the variable states. The optimal decision is derived by verifying the concerned NFRs and reducing the decision space. Our approach is implemented through the demonstration of a mobile information system.

## 1. Introduction

The self-adaptive system (SAS) is a novel computing paradigm in which the software is capable of adjusting its behavior in response to meaningful changes in the environment and itself [1]. The ability of adaptation is characterized by self-* properties, including self-healing, self-configuration, self-optimizing and self-protecting [2]. Innovative technologies and methodologies inspired by these characteristics have already created avenues for many promising applications, such as mobile computing, ambient intelligence, ubiquitous computing, etc.

Modern software systems interact with other systems, devices, sensors and people intensively. Such an operational environment may be inherently changeable, which makes self-adaptiveness become an essential feature. Context can be defined as the reification of the environment [3] that is whatever provides as a surrounding of a system at a time. It provides a manageable and manipulable description of the environment. Context is essential for the deployment of self-adaptive software. As the environment is changeable, the context is unstable and ever changing and the system is desired to perform different behaviors accordingly.

Requirements Engineering (RE) for SASs primarily aims to identify adaptive requirements, specify adaptation logic and build adaptation mechanisms [4]. At this stage, both functional requirements (FRs) and non-functional requirements (NFRs) [5] are captured within well-formed requirements models. Conducting context analysis at requirements phase will be worthwhile at the design and development phases, because contexts may influence the decisions about what to build and how to build them.

Research challenges of RE for SASs are composed of several aspects, including modeling adaptation mechanism, dealing with uncertainty, achieving adaptation and requirements verification [4]. Many research works in the literature have shown remarkable progress in providing solutions to these challenges. For modeling adaptation mechanism, Qureshi et al. [6] provide a requirements modeling language with diagrammatic syntax for SASs. Cheng et al. [7] introduce a goal-based modeling approach to develop the requirements for dynamically adaptive systems, while explicitly factoring uncertainty into the process and resulting requirements. For dealing with requirements uncertainty, a research agenda is provided in [8]. The author argues requirements for self-adaptive systems should be viewed as runtime entities that can be reasoned over in order to understand the extent to which they are being satisfied and to support adaptation decisions that can take advantage of the systems' self-adaptive machinery. FLAGS [9] is proposed for mitigating the requirements uncertainty by extending the goal model with adaptive and fuzzy goals. RELAX [10] is a formal requirements specification language, which is defined in terms of temporal fuzzy logic, describing relaxed requirements and the environment. For achieving adaptation, FUSION [11] uses online learning to mitigate the contextual changes in context and reconfigure features from feature pool to unanticipated changes. The method utilizes linear equations to simulate the system for reducing time complexity of decision making. POISED [12] improves the quality attributes of a software system through reconfiguration of components to achieve a global optimal configuration for the software system. Besides, some other approaches to achieving adaptation are based on reasoning with goal model [13, 14]. For requirements verification, Goldsby et al. [15] provide AMOEBA-RT, a run-time monitoring and verification technique that provides assurance that dynamically adaptive software satisfies its requirements. Filieri et al. [16] provide a model checking based approach for verification, in which reliability models are given in terms of Discrete Time Markov Chains which are verified against a set of requirements expressed as logical formulae. Epifani et al. [17] lay the foundations for an iterative model-driven development, which aims at verifying that an implementation satisfies non-functional requirements. In their approach, if the resulting running system behaves differently from the assumptions made at design time, the feedback to the model shows why it does not satisfy the requirements and lead to a further development iteration. Ghezzi et al. [18] put forward the proposal of quantitative verification at runtime for self-adaptive service-based software. The same author proposes a model-based approach that enables software engineers to assess their design solutions for software product lines in the early stages of development [19].

During implementing self-adaptation, two concerns that are lacking in discussing should be put forward: reliability and performance. Firstly, as discussed earlier, contexts are intrinsically changing which may cause the violation of the system-to-be. For example, in mobile computing systems domain, the precision of locating with GSM is highly depended on the status of network. The higher bandwidth is, the more precise location the system will get. In other words, the low bandwidth will affect the reliability of GSM. Hence, self-adaptive systems need to be built with the mechanism of monitoring contextual changes, analyzing violations of NFRs and making adaptation decision. Secondly, performance of SASs is another essential concern for SASs, especially when systems are supposed to be deployed in the time/resource-limited environment. System performance comes with several dimensions, such as utility, time costs and energy costs. Specifically, we use utility to describe the convenience of system behaviors for users. For example, in mobile computing domain, the utility of searching by voice is higher than searching by text, because the interaction between system and users in the former scenario makes searching more convenient. Time costs intuitively refer to the time consumed by performing certain system behavior, e.g. monitoring contexts and searching the data base, while energy costs describe the energy consumed by these tasks. Thus, when implementing the adaptation, we should also consider these performance dimensions.

In this paper, we provide a model-driven approach to achieving self-adaptation from the requirements engineering perspective. Modeling is considered as an important conceptual tool in RE stage of software development. Requirements models are built to better understand and reason about the qualities a system should exhibit in order to fulfill its goals. They are also used to support systematic interaction with the stakeholders in requirements elicitation and to help crystallize design decisions and evaluate the trade-offs among them. To modeling adaptation mechanisms, we build adaptation goal model by extending traditional goal model with new modeling elements representing the monitor-analyze-plan-execute process, which is known as MAPE control loop [20]. Then, the reliability and performance are specified as tags to each system task. Particularly, to formally describe reliability of SASs, we consider the uncertainty of system tasks, which can be quantitatively expressed with the failure rate of each task. Different probability distribution will lead to different system reliability. Thus, we choose to use Bayesian methods for reasoning with these probabilities and performance constraints. To this end, we investigate the system behavior from the requirements model by applying several operators adopted from Communicating Sequential Processes (CSP). With the well-formed transformation relations and algorithm, the requirements model can be expressed with the corresponding processes and behavior model, which displays as a Labeled Transition System (LTS). We use the Process Analysis Toolkit (PAT) to verify the properties of the generated process-based LTS and the results imply the reasonability and consistency of the transformation process. Thereafter, the tagged adaptation goal model and the LTS are integrated into probabilistic Markov models, i.e. the Discrete-Time Markov Chains (DTMCs) and Continuous-Time Markov Chains (CTMCs). The non-functional properties are specified with Probabilistic Computation Tree Logic (PCTL). We propose to achieve structural adaptation through verifying PCTL properties with DTMCs and to derive parametric adaptation through verifying Reward properties with CTMCs. The verification processes are implemented through model checking within PRISM. An application from the mobile computing domain is leveraged as illustration throughout the approach.

Our contributions are multifold. First is that we present a general method for describing adaptation mechanisms within requirements models of SASs. Compared with the related work, our model is more concise but easier to be manipulated. The well-formed structure provides the possibility of generating system behavior model. Second, we propose a flexible notation for specifying the reliability and performance of SASs, which are uncertain and varying according to related contexts. These notations can help characterize adaptation behaviors. Third, the approach provides processes for transforming requirements model to corresponding behavior model, which can be applied to SASs of any other domains that are represented with goal models. It can be leveraged as the basis for behavior analysis in following stages of software development. Fourth, the approach presents a step-by-step way to deriving adaptation decisions among alternative candidates. Requirements are consid-ered as runtime properties and the verification-based adaptation is conducted based on strict mathematical foundations. Last but not least, the whole approach is driven by requirements models and behavior models. Thus, the approach bridges the gap between static and dynamic attributes of SASs and the gap between adaptation logic and adaptation behavior.

The paper is structured as follows. Section 2 illustrates the motivating example and the overall approach, in order to deliver an easy understanding of this paper. Section 3 introduces the overview of the proposed approach. The concepts and modeling notations of adaptation goal model are presented in Section 4. Section 5 provides the tagged adaptation goal model by specifying reliability and performance characteristics. Section 6 presents the sequential adaptation goal model with CSP operators and transformation to the behavior model. Reliability-related behavior model (DTMCs) is presented in Section 7, while Performance-related behavior model (CTMCs) is described in Section 8. Section 9 provides the evaluation of our approach with the motivating example. Related works are discussed in Section 10, followed by conclusion and future work in Section 11.

## 2. Motivating Example

To illustrate the proposed approach, we consider adopting the push notification technology from the mobile computing domain. Typically, pushing notifications is a technique used by apps to alert smartphone owners on content updates, messages, and other events that users may want to be aware of. This technique has been successfully developed as API on iOS systems, such as Prowl (http://www.prowlapp.com) and Pushover (https://pushover.net/). These applications focus on receiving the needed information timely, no matter where the user is. However, in the location-related scenarios, the application's performance tightly related to the user's location, e.g. learning application on smartphone [21, 22].

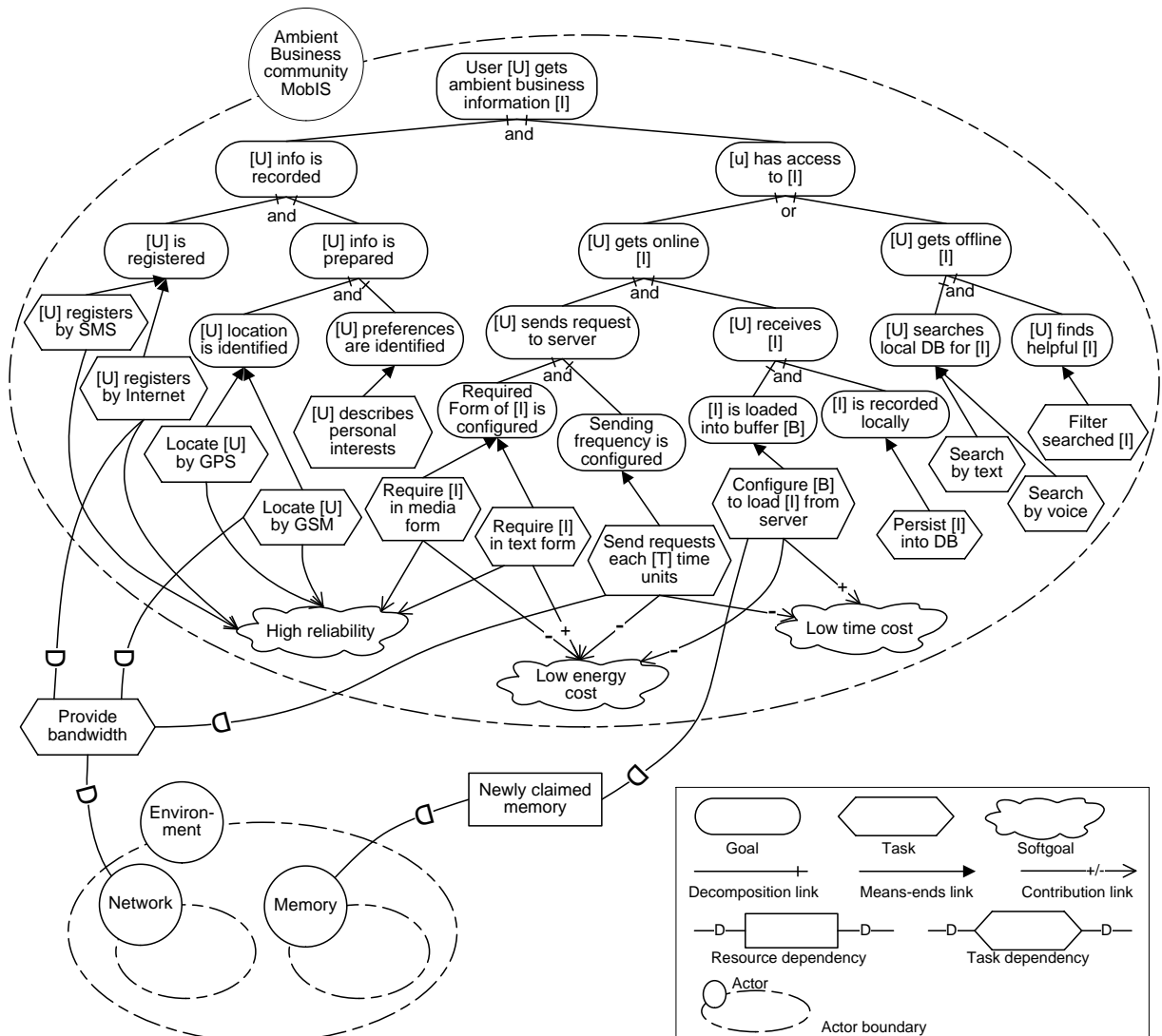

**Figure 1 Goal Model of Ambient business community MobIS**

The ambient business community Mobile Information System (MobIS) is the software deployed on smartphones to support pushing business information to nearest customers in a smart business community. The objective of pushing business information is to notify users of surrounding information and events, such as the goods on sale, goods' description, comments of customers, etc. Figure 1 presents a requirements model built with strategic rationale (SR) diagram in i* framework [23], which is defined to model and reason about both the system and its organizational environment.

The top goal of MobIS is *user gets the ambient business information*, which can be decomposed into two subgoals with Decomposition-link: *user info is recorded* and *user has access to business info*. The Decomposition-links describe both AND-relation and OR-relation, which are displayed as and/or notations. *user info is recorded* is further decomposed into *user is registered* and *user info is prepared*. The goal *user is registered* can be achieved through two tasks: *register by SMS* and *register by Internet*. The Means-Ends links are used to provide alternative operational tasks for the target goal. User information consists of two parts: location information and preferences information. The former one aims to recognize user's location (by GPS or GSM) while the later one identifies personal interests. Users can access business information in either *online* mode or *offline* mode. For *online* mode, MobIS *sends request timely* for information of either *media form* or *text form*. To *receive information*, MobIS needs to *load data into buffer* and then data *persists into database*. For *offline* mode, user can search local database either by *voice* or *text*. Besides, user can *filter* the retrieved information. Four softgoals that capture NFRs need to be concerned at runtime: *High reliability*, *Low energy costs*, *Low time costs* and *High Utility*. Actually, each task has different contribution to these softgoals. The contribution-links in Figure 1 are partial contribution relations. The quantitative descriptions of these contributions and instantiations of NFRs will be provided in Section 7 and Section 8.

Two environment actors are considered in this example. *Network* describes the external environment that changes automatically, with the task of *show bandwidth to sensors* of outside systems. *Memory* is the internal environment whose value can be tuned according to adaptation decisions, with the task of *prepare spared memory* for self-adaptation. MobIS needs network to *provide bandwidth*. This relation is described by Task-dependency. While MobIS needs *claims memory* from memory pool. It is presented as Resource-dependency.

## 3. Approach Overview

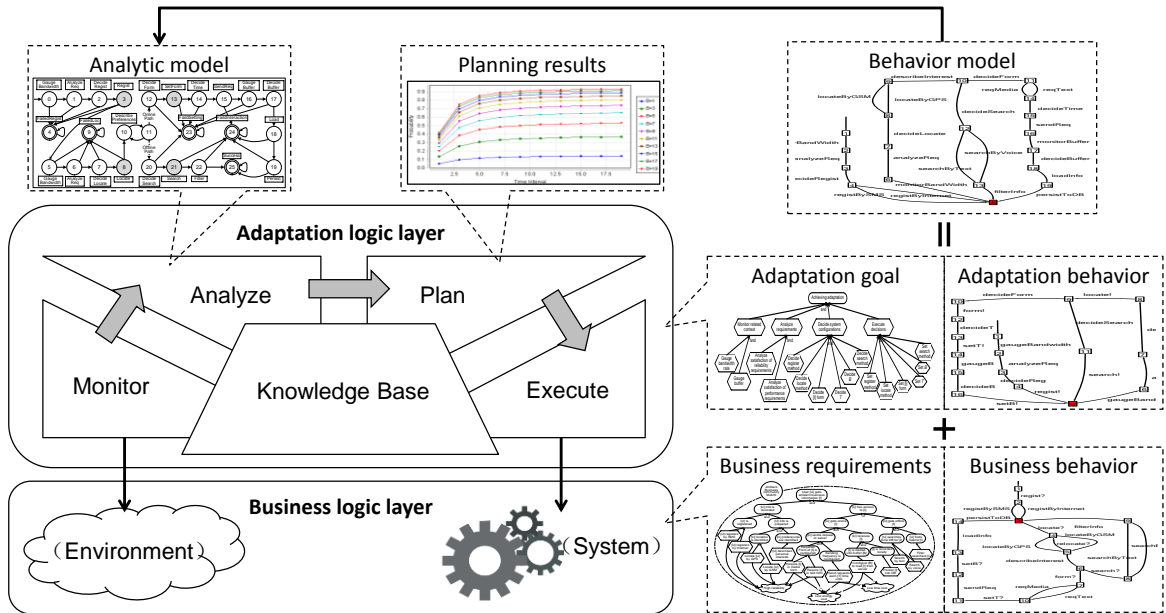

**Figure 2 Generating relations between models in our model-driven approach**

Models are the core concepts and artifacts in our model-driven approach. The approach contains several steps and each step generates the corresponding model. The generating relations between models are described in Figure 2. Models in the line-blocks capture static characteristics of SASs while models in the dash-blocks describe dynamic characteristics of SASs. The concrete steps are stated as follows.

***Building Adaptation Goal Model (AGM)*** is the initial step of our approach. Traditional goal model only capture business logic (Figure 1), while goal model of SASs needs to be endowed with adaptation logic. We build the adaptation mechanism adopted from MAPE loop by adding incremental tasks to traditional tasks that need to be reconfigured at runtime. In this step, we consider both modeling structural adaptation logic and parametric adaptation logic.

***Building Tagged Adaptation Goal Model (TAGM)*** is conducted based on the completion of AGM. In this step, tags of adaptation concerns, i.e. reliability and performance, are described within AGM. These tags will be used to compute the adaptation decisions at runtime.

***Building Sequential Adaptation Goal Model (SAGM)*** focuses on providing an efficient way to transforming goal model to corresponding behavior. SAGM is generated by adding several sequential operators to AGM, which are adopted from CPS. By this means, we explore the transformation patterns of different decomposition patterns of goal model. These transformation relations are used as the basis of deriving behavior model.

***Transforming to Behavior Model*** aims at transforming the system goal model to system behavior by using the above transformation patterns. The derived behavior model is expressed as a Labeled Transition System (LTS). Business logic can be verified with the generated LTS, which confirms the reasonability and consistency of the transformation process.

***Generating Reliability-related Model*** integrated reliability tags of TAGM with LTS. The resulting model is expressed as a variable DTMC, which can be used for Bayesian reasoning with the transition probabilities.

***Generating Performance- related Model*** merges performance tags of TAGM with LTS. This step is completed by producing a Rewarded variable CTMC, which can be used for Bayesian reasoning with the transition rate.

***Runtime verification*** is the last step of our approach. Reliability requirements are expressed as PCTL properties, which can be verified within the DTMC. Performance requirements are described as CSL properties, which can be verified within the CTCM.

## 4. Modeling Adaptation Logic

This section presents the concept and modeling elements of AGM. Then we describe three kinds of adaption scenarios and discuss how to model adaptation mechanisms for each scenario based on the proposed modeling notations.

### 4.1 Adaptation Goal Model

**Definition 1 (Adaptation Goal)** Adaptation goals refer to the goals that are related to adaptation requirements and should be achieved with adaptation tasks.

**Definition 2 (Adaption Task)** Adaptation tasks refer to the tasks that should be implemented through making adaptation decision.

**Definition 3 (Adaptation Goal Model)** Adaptation Goal Model is defined as
$$AGM = (G, AG, SG, T, AT, M, A, P, E, R_M, R_D, R_C)$$
where $G$ is the set of ordinary goals, $AG$ is the set of adaptation goals, $SG$ refers to the set of softgoals, $T$ refers to the set of ordinary tasks, $AT$ refers to the set of adaptive tasks. $M$, $A$, $P$ and $E$ consists of the tasks that perform monitoring, analyzing, planning and executing correspondingly, which characterize the adaptation mechanisms. $R_M: G \times T \cup AG \times AT$ is the extended Means-Ends relation. $R_D: G \times G \cup T \times T \cup AT \times \{M, A, P, E\}$ is the extended Decomposition relation. $R_C: T \times SG \cup \{M, A, P, E\} \times SG$ is the extended Contribution relation. These modeling elements and relations are comprised in the metamodel presented in Figure 3.

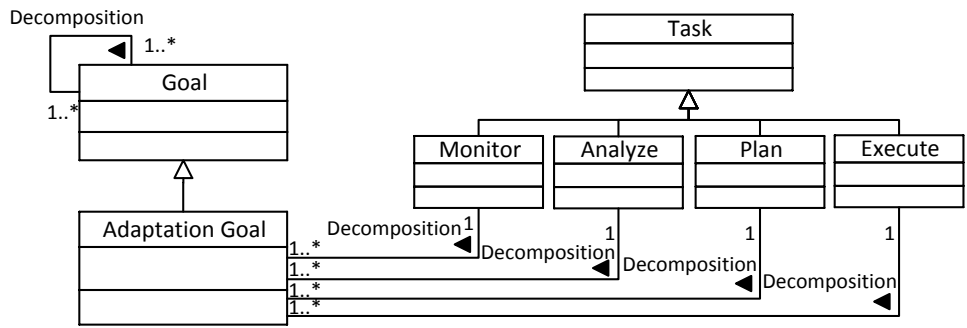

**Figure 3 Metamodel of adaptation goal model**

### 4.2 Adaptation Mechanism

Based on the modeling elements proposed above, we discuss how to modeling adaptation logic and mechanisms within the requirements model. To this end, we first describe the adaptation scenarios considered in our approach, then for each scenario, we model the related adaptation mechanism.

Recall the self-* properties introduced in Section 1, which includes self-healing, self-configuration, self-optimizing and self-protecting. In our approach, we consider the properties of self-healing, self-configuration and self-optimizing. For self-protecting, readers can refer to [24] for more details about self-protecting software systems. For each of the former three self-* properties, we explore the adaptation scenario (AS) related with reliability and performance. Table 1 shows the causes, effects and countermeasures in each scenario. Notice that AS1 describes the structural adaptation while AS2 describes the par-ametric adaptation. Differences of the two are introduced in [25]. However, in this paper, structural adaptation is discussed in a broader sense, which means the adaptation is implemented based on alternative task structures, either by configuring the specific components, e.g. GPS or GSM, or by configuring the optional tasks, e.g. require information in media form or text form. Another characteristic that should be noticed is that for an optimal adaptation decision, self-configuration and self-optimizing are achieved simultaneously.

For each scenario, an adaptation goal and corresponding adaptation mechanisms should be figured out in the AGM. In our BobIS example, the modeling results are presented in Figure 4. $AG_1$ is the adaptation goal for AS1, which is generated from the original goal *user location is identified*. $AG_1$ can be achieved by implementing $AT_1$, which is further decomposed with MAPE tasks. $AG_2$ is a newly refined adaptation goal, considering the failure of locating task. Once a failure is detected, $AT_1$ should be invoked to derive the correct location. Different from the above two adaptation goals, $AG_3$ and $AG_4$ are achieved through parametric adaptation. For $AG_3$, the sending time interval is affected by the network. When the bandwidth is high, time interval is set to a small value for getting newer information. In this situation, buffer needs to be set to an appropriate value for loading information timely. The larger buffer value is, the faster system processes the load data. With these modeling elements, a traditional goal model is converted to the adaptation goal model. The specifications of AGM and general algorithms of MAPE mechanisms are described in our primary work [26].

**Table 1 Adaptation scenarios for different self-* properties**

| AS | Self-* property | Cause | Effect | Countermeasure | Example |
|---|---|---|---|---|---|
| AS1 | self-configuration, self-optimizing (structural) | Contextual changes | Decrease system reliability and increase operational costs | Monitor contextual changes and generate the optimal configurations of system structures | $AG_1$ |
| AS2 | Self-healing | Task failures | Lost connection with following tasks and decrease system reliability | Monitor the implementation of related tasks and redo the task if failure is detected | $AG_2$ |
| AS3 | self-configuration, self-optimizing (parametric) | Contextual changes | Decrease system reliability and increase operational costs | Monitor contextual changes and generate the optimal configurations of related parameters | $AG_3$, $AG_4$ |

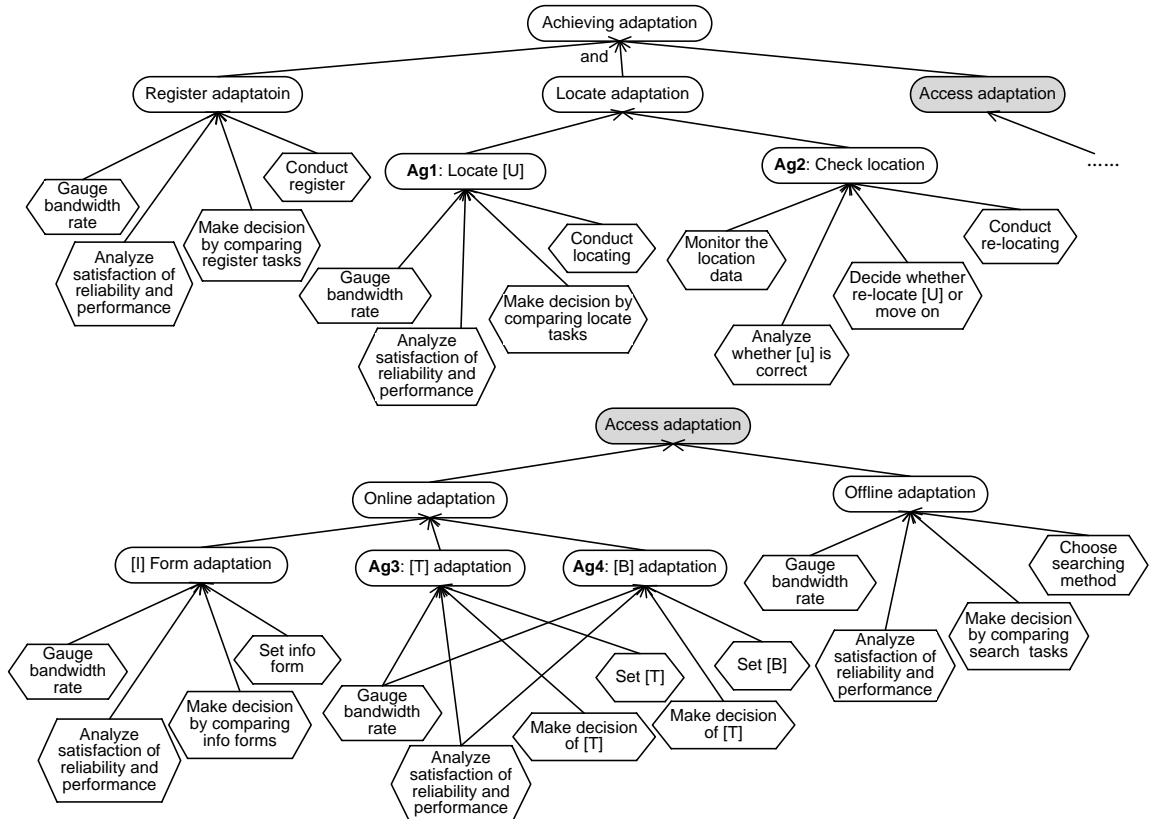

**Figure 4 Modeling of adaptation goals and adaptation mechanisms.**

## 5. Specifying Reliability and Performance

In adaptation goal model, each goal is refined to operational tasks that can be performed by SASs. However, the reliability and performances are varying according to contexts at runtime, as we discussed in Section 1. This section presents the specifications of reliability and performance in AGM. With these specifications, sequence diagrams of SASs are used to display the performance of adaptation processes.

### 5.1 Tagged Adaptation Goal Model

**Definition 4 (Tagged Adaptation Goal Model)** A Tagged Adaptation Goal Model (TAGM) is generated by extending the tasks of AGM with reliability and performances tags. It is denoted as
$$TAGM = (G, AG, SG, \{T, TAG\}, AT, \{M, TAG\}, \{A, TAG\}, \{P, TAG\}, \{E, TAG\}, R_M, R_D, R_C).$$
For a certain task $t$ with parameter $p$, $TAG(t, p) \coloneqq \{c, FP(c), U(c), TC(c, p), EC(c, p)\}$, where $c$ refers to the context value, $FP(c)$ is the failure probability of task in context $c$, $U(c)$ is the utility of task in context $c$, $TC(c, p)$ is the time costs with parameter $p$, $EC(c, p)$ is the energy costs with parameter $p$. For structural tasks, $TC(c, p)$ and $EC(c, p)$ are simplified as $TC(c)$ and $EC(c)$. For a continuous system, $FP$, $U$, $TC$ and $EC$ are the functions of context $c$. It is hard to identify the mathematic formulae between SASs and contexts [27]. However, for a discrete system, we can define the values of tags according to contextual classification. In RE stage, a well-conducted contextual modeling and analysis approach is proposed in [28]. Figure 5 present the modeling of tags in TAGM of MobIS, according to the classification of network status.

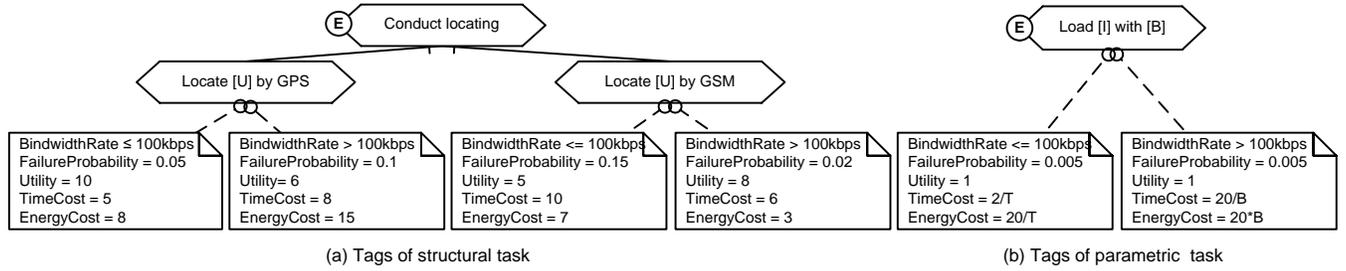

**Figure 5 Modeling of tags with contextual classification.**

### 5.2 Tagged Adaptation Sequence Diagram

Sequence Diagram is an effective model for describing the dynamic interaction between objects and displaying the behaviors of the system-to-be. Hence, for modeling SASs, the processes of adaptation mechanisms can be clearly figured out within the sequence diagram.

**Definition 5 (Tagged Adaptation Sequence Diagram)** A Tagged Adaptation Sequence Diagram (TASD) describes the interaction between MAPE processes. It is defined as $TASD = (O, \{MES, TAG\}, FRA)$, where $O$ in the object set, $MES$ is the set of messages, TAG is the tags related to the messages, FRA is the set of fragments including *loop* and *alt*. To describing adaptation process, the object is denoted as $O = (AT, M, A, P, E, C, R)$, where $C$ is the set of contexts related to the $AT$, $R$ is the related resources.

Figure 6 presents the TASD of adaptation task $AT_1$. The *alt* fragment expresses the messages sent according to different location decisions. Figure 7 presents the TASD of adaptation task $AT_3$. The loop fragment indicates requires are sent after each time interval.

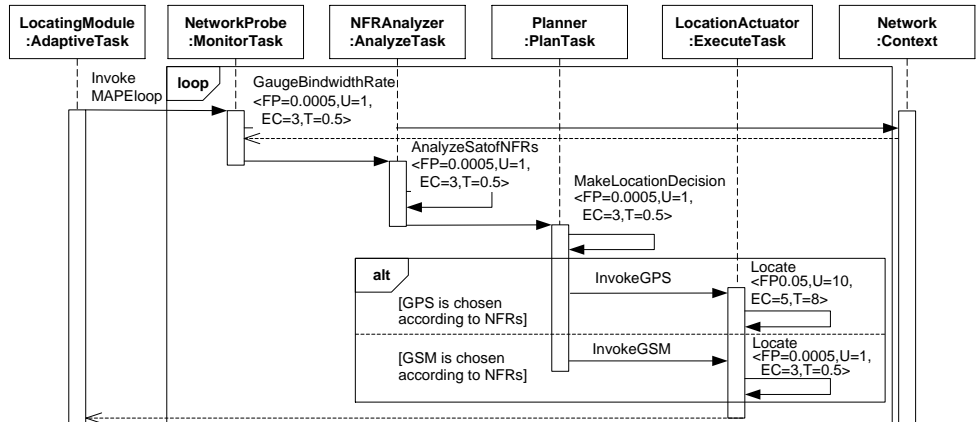

**Figure 6 Tagged Adaptation Sequence Diagram of $AT_1$**

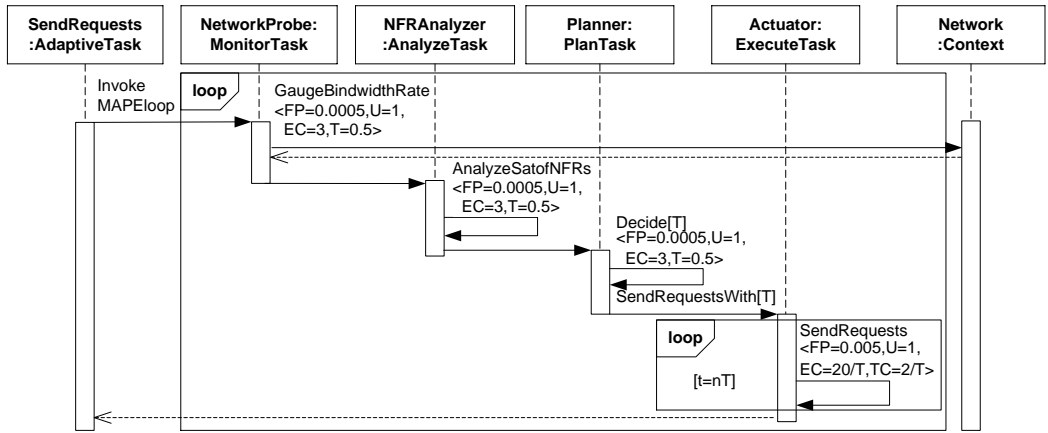

Figure 7 Tagged Adaptation Sequence Diagram of $AT_3$

## 6. Modeling Adaptation Behavior

Though sequence diagram can express the dynamic interaction processes of adaptation mechanisms, it is hard to analyze dynamic behavior at runtime for the lack of formalization. To efficiently formalize the behavior of adaptation mechanisms and the behavior of system, this section provides a CSP-based modeling method that transforms the requirements model to the corresponding behavior model.

### 6.1 Sequential Adaptation Goal Model

**Definition 7 (Sequential Adaptation goal Model)** A Sequential Adaptation Goal Model (SAGM) is defined as
$$SAGM = (GP, TP, A, \alpha TP, Traces(TP), Chan, Op, \Delta)$$
-*GP* denotes the set of goal processes that are conducted for achieving goal *G*. *GP* is specified by what SASs can do.
-*TP* denotes the set of task processes that are performed by task *T*. *TP* is defined by the system behavior.
- *A* denotes the set of events (tasks) engaged in *TP*. Notice that $\tau \in A$ is the hidden event.
- $\alpha TP$ denotes the alphabet of *TP*, which contains the events possibly engaged in *TP*, e.g. in Figure 4(a),
$\alpha ATP_1 = \{GaugeBandwithRate, AnalyzeSatofNFRs, MakeLocationDecision, LocateByGPS, LocateByGSM\}$
- $Traces(TP)$ is a finite sequence of events (tasks). A *TP* can be specified by the set of traces denoting its possible behavior, e.g. the traces of $AT_1$ are:
$$\langle GaugeBandwithRate, AnalyzeSatofNFRs, MakeLocationDecision, LocateByGPS \rangle$$
$$\langle GaugeBandwithRate, AnalyzeSatofNFR, MakeLocationDecision, LocateByGSM \rangle$$
-*Chan* is the set of channel between processes. The message set on channel of *TP* is defined as $\alpha C(TP) = \{v | chan.v \in \alpha TP\}$, where *chan* is the name of channel and *v* is the value of the delivered message.
-*Op* is the set of operators, including: *Prefix* (→), *Sequential Composition* (;), *Parallel Composition* (||) and *Choice* (□).

- →: A process which may participate in task *t* then act according to TP is written as $t \rightarrow TP$.
- ; : The sequential composition of GP1 and GP2, written GP1; GP2, acts as GP1 until GP1 terminates and then proceeds to act as GP2.
- || : The parallel composition of GP1 and GP2, written GP1||GP2, acts both GP1 and GP2 concurrently.
- □: Choice operator allows system a choice of behavior according to what tasks are conducted. The process $(t1 \rightarrow TP1) \square (t2 \rightarrow TP2)$ acts as TP1 if *t1* occurs, while acts as TP2 if *t2* occurs.

-$\Delta \subseteq R_M \times Op \cup R_D \times Op$ is the mapping relation. $R_M$ and $R_D$ have the same meaning within definition 3. This relation is detailed in Table 2.

Summarly, in SAGM, goals are mapped to goal process; tasks are mapped to task process and tasks events (only for leaf-node tasks); AND-decompositions are mapped to Sequential and Parallel operators; OR-decompositions and Means-Ends links are mapped to Choice operators; Dependency relations are mapped to communication between task processes.

To generate the behavior of SASs based on SAGM, we need to further investigate the usages of the operators. By investigating the relations of goals and tasks in the goal model presented in Figure 1, we split these relations into some decomposition and dependency patterns. Each pattern can be transformed to process behavior with the proposed operators. Table 2 elaborates the relation patterns, which can be specified with proposition logic in the goal model. After appending the process operators, relation patterns are specified with CSP descriptions that can be used to derive corresponding behavior models, presented as Labeled Transition Systems (LTSs). Notice that, in behavior patterns, *tau* is the hidden event of related processes.

A LTS is defined as LTS = $(S, s_0, \Sigma, \rightarrow_s)$, where $S$ is a finite set of states, $s_0 \in S$ is the initial state, $\Sigma$ is a finite set of actions, $\rightarrow_s \subseteq S \times \Sigma \times S$ is a transition relation. The behavior patterns in Table 2 imply that a transition represents a certain system task while a state refers to completion of the task.

**Table 2 CSP descriptions and behavior of relation patterns**

| Relation pattern | Proposition logic | Operator | CSP Specifications | Behavior pattern |
|---|---|---|---|---|
| g1 and g2, g3 | $g2 \wedge g3 \rightarrow g1$ | ; | GP1=GP2;GP3<br>GP2=tau1->tau2->GP3<br>GP3=tau3->GP2 | tau1, tau2, tau3 |
| g1 and g2, g3 | $g2 \wedge g3 \rightarrow g1$ | \|\| | GP1=GP2\|\|GP3<br>GP2=tau1->tau2->GP2<br>GP3=tau1->tau3->GP3 | tau1, tau2, tau3 |
| g1 or g2, g3 | $g2 \vee g3 \rightarrow g1$ | □ | GP1=GP2[]GP3<br>GP2=tau1->tau2->GP1<br>GP3=tau1->tau3->GP1 | tau1, tau2, tau3 |
| t1 and t2, t3 | $t2 \wedge t3 \rightarrow t1$ | ; | TP1=TP2;TP3<br>TP2=tau1->t->TP3<br>TP3=t3->TP2 | tau1, t2, t3 |
| t1 and t2, t3 | $t2 \wedge t3 \rightarrow t1$ | \|\| | TP1=TP2\|\|TP3<br>TP2=tau1->t2->TP2<br>TP3=tau1->t3->TP3 | tau1, t2, t3 |
| t1 or t2, t3 | $t2 \vee t3 \rightarrow t1$ | □ | TP1=TP2[]TP3<br>TP2=tau1->t2->TP1<br>TP3=tau1->t3->TP1 | tau1, t2, t3 |
| g1, t1, t2 | $t2 \vee t3 \rightarrow g1$ | □ | GP1=TP2[]TP2<br>TP2=tau1->t1->GP1<br>TP2=tau1->t2->GP1 | tau1, t1, t2 |
| A(t1)-D-B(t2) | $t2 \rightarrow t1$ | \|\| | PAR=A\|\|B<br>A=c?0->t2->A<br>B=t1->c!0->B | t1, t2, c.0 |

## 6.2 Behavior Model

### 6.2.1 Behavior of Adaption Mechanisms

According to the above CSP descriptions and behavior patterns, we can derive the behavior patterns of structural adaptation and parametric adaptation, which are provided in Figure 8. It shows that structural adaptation has alternative behavioral structures for executing the adaptation decision, while parametric adaptation has only one structure with alternative parameters.

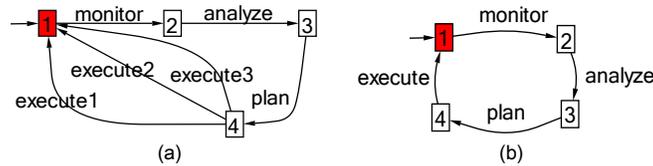

**Figure 8 Behavior patterns of structural adaptation (a) and parametric adaptation (b)**

Based on these two patterns of MAPE processes, we can generate the behavior models of adaptation mechanisms in our MobIS example (Figure 4). Figure 10(a) presents the behavior for achieving *AG1*. Two alternative behaviors are *locateByGPS* and *locateByGSM*, which have been modeled as executing tasks in the AGM. Figure 10(b) shows the behavior for achieving *AG2*. Actually, it contains both the adaptation loops of AG1 and AG2. The right loop acts when the location is accurate, while the left loop acts when the locating task fails. Once the locating failure is detected, AG1 is invoked again to relocate users. Figure 10(c) describes the adaptation behavior for AG3. Notice that the self-loop in the LTS is produced by the status of related task. For instance, as depicted in Figure 7, the task *sendReq* is performed each *T* time unit. If there is no need to reconfigure the value of *T*, *sendReq* will be continuously performed. Hence, the behavior is denoted as a self-loop transition. Figure 10(d) depicts the behavior for AG4. Similarly, the task *loadInfor* is also a self-loop transition.

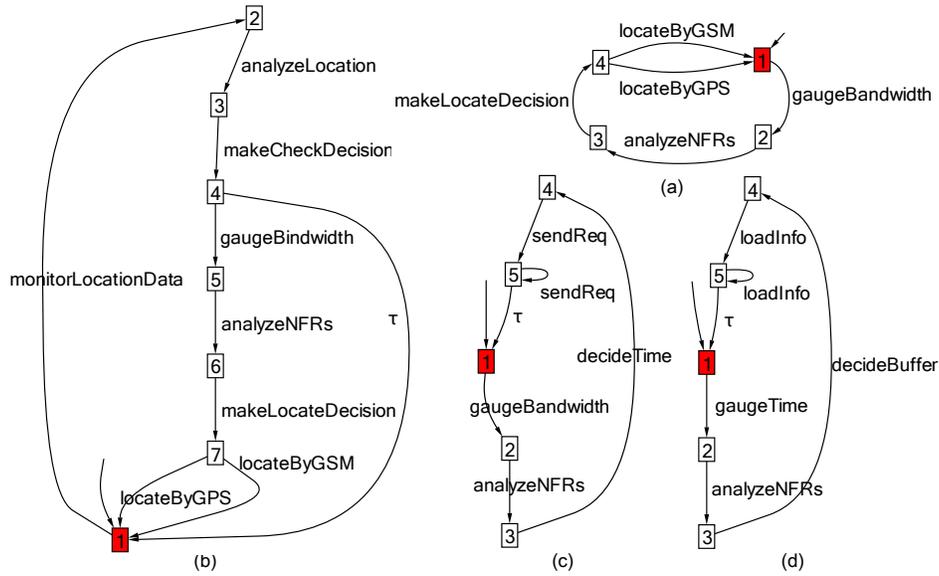

**Figure 9 Behaviors for AG1 (a), AG2 (b), AG3 (c) and AG4 (d) of MobIS example**

### 6.2.2 Behavior of Self-Adaptive Systems

This subsection presents the transformation from SAGM of a SAS to its behavior model. Firstly, we need to derive the CSP specifications of SASs. Figure 10 presents the specifications of the BobIS example based on the idea of iterative Depth-First Search (DFS). It is produced by a top-down process. The process operators imply the mapping from $R_M, R_D$ and $R_C$ of AGM to *Op* of SAGM. For example, "MobIS=RecordU;UAccessI" implies the AND-decomposition of the root goal is mapped to the sequential composition operator. "UAccessI=GetOnline[]GetOffline" implies the OR-decomposition of the goal *user has access to info* is mapped to the choice operator.

```
//CSP Specification of SAGM of MoBIS
//decomposition of root node
MobIS = RecordU;UAccessI;
    //left child node
    RecordU = RegistU;PrepareU;
        //left subtree
        RegistU = monitorBandWidth->analyzeNFRs->makeDecision->(R1[]R2);
            R1 = registBySMS->PrepareU;
            R2 = registByInternet->PrepareU;
        //right subtree
        PrepareU = IdentifyLocation;IdentifyPrefer;
            IdentifyLocation = Locate;CheckLocation;
                Locate = monitorBandWidth->analyzeNFRs->makeDecision->L1[]L2;
                    L1 = locateByGPS->CheckLocation;
                    L2 = locateByGSM->CheckLocation;
                CheckLocation = monitorLocationData->analyzeLocation->
                            makeCheckDecision->(IdentifyPrefer[]IdentifyLocation);
            IdentifyPrefer = describeInterest->UAccessI;
    //right child node
    UAccessI = GetOnline[]GetOffline;
        //left subtree
        GetOnline = SendReq;ReceiveI;
            SendReq = ConfigForm;ConfigTime;
                ConfigForm = F1[]F2;
                    F1 = reqMedia->ConfigTime;
                    F2 = reqText->ConfigTime;
                ConfigTime = decideTime->Send;
                    Send = sendReq->(Send[]ReceiveI);
                        ReceiveI = LoadInfo;PersistInfo;
                            LoadInfo = decideBuffer->Load;
                                Load = loadInfo->(Load[]PersistInfo);
                            PersistInfo = persistToDB->PrepareU;
        //right subtree
        GetOffline = SearchDB;FindInfo;
            SearchDB = S1[]S2;
                S1 = searchByText->FindInfo;
                S2 = searchByVoice->FindInfo;
            FindInfo = filterInfo->PrepareU;
```

**Figure 10 CSP description of MobIS**

To check whether these specifications correctly describe the business logic and adaptation logic of the MobIS, we apply the Process Analysis Toolkit (PAT) to the simulation and verification of the above specifications. After simulation, the generated LTS is depicted in Figure 11. It merges four loops of adaptation mechanisms presented Figure 9. With the LTS, we can verify several properties of the system.

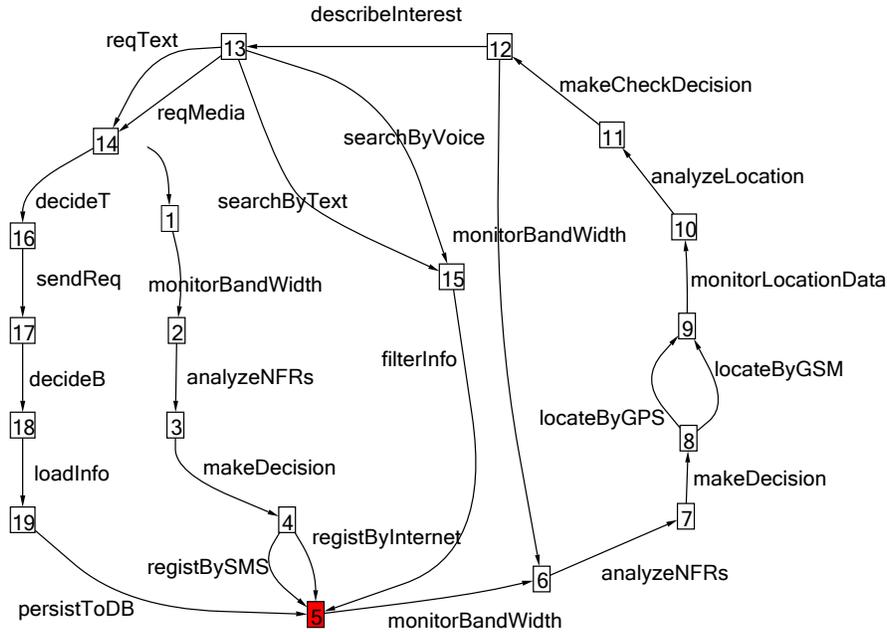

**Figure 11 Generated LTS from CPS specifications of MobIS**

Four kinds of properties of the LTS are verified, as presented in Figure 12. Safety properties capture primary demands of systems behavior. *Deadlock-free* depicts the system has no deadlock state. N*on-terminating* depicts the system has no terminated states. *Divergence-free* refers to system behavior can be recursively controlled. *Deterministic* property means for any system state, there is no two out-going transitions leading to different states but with the same task. Refinement properties describe the decomposition relations in AGM. Liveness and reachability properties are defined based on Linear Temporal Logic (LTL). Liveness properties represent the properties that should always be satisfied. The liveness assertion in Figure 12 means that once the process *IdentifyLocation* acts, the event *monitorLocationData* can always be engaged. Reachability properties capture the states that can be finally reached in a certain behavior path. The assertion in Figure 12 depicts that once online mode is chosen, MobIS can finally persist information to database. While, once offline mode is chosen, user can finally filter the searched information. The correctness of these verified properties implies the reasonability of the transformed LTS and the consistency of it with business logic and adaptation logic.

```
//safety properties
#assert MoBIS deadlockfree;
#assert MoBIS nonterminating;
#assert MoBIS divergencefree;
#assert MoBIS deterministic;
//refinement properties
#assert GetOnline refines UAccessI;
#assert GetOffline refines UAccessI;
//liveness properties
#assert IdentifyLocation |= []<>monitorLocationData;
//reachability properties
#assert GetOnline |= <>persistToDB;
#assert GetOffline|= <>filterInfo;
```

→ Verification →

| | Assertions |
|---|---|
| ✓ 1 | MoBIS() deadlockfree |
| ✓ 2 | MoBIS() divergencefree |
| ✓ 3 | MoBIS() nonterminating |
| ✓ 4 | MoBIS() deterministic |
| ✓ 5 | GetOnline() refines UAccessI() |
| ✓ 6 | GetOffline() refines UAccessI() |
| ✓ 7 | IdentifyLocation() |= []<> monitorLocationData |
| ✓ 8 | GetOnline() |= <> persistToDB |
| ✓ 9 | GetOffline() |= <> filterInfo |

**Figure 12 Properties and verification results**

## 7. Reliability-related Adaptation

Based on the above extended requirements models and behavior models, we elaborate how these models can be used for deriving runtime adaptation decision. We divide the adaptation process into two phases of adaptation. The first phase of adapta-tion is performed according to reliability requirements, while the second step is conducted according to performance requirements. This section presents the reliability-based adaptation.

## 7.1 Probabilistic Behavior Model

Discrete Time Markov Chains (DTMCs) are a widely accepted formalism to model reliability of systems. In our work, these components are represented with system tasks that are conducted by related components. Particularly, DTMCs are proved to be useful for predicting reliability [30]. The adoption of DTMCs implies that the modeled system meets the Markov property[31]. DTMCs are discrete stochastic processes with the Markov property, according to which the probability distribution of future states depends only upon the current state. They are defined as a Kripke structure with probabilistic transitions among states. States represent possible configurations of the system. Transitions among states occur at discrete time and have an associated probability. Traditional DTMCs have fixed structures, which is not appropriate for SASs. Hence, we extend DTMCs with a set of *variable states*. Formally, a Variable DTMC is a tuple $(S_I, S_v, s_0, \boldsymbol{P}, L)$ where

- $S_I$ is a finite set of invariable states.
- $S_v$ is a finite set of variable states.
- $s_0 \in S$ is the initial state
- $\boldsymbol{P}: (S_I \cup S_v) \times S_I \to [0,1]$ is a set of transition probability matrix, where $\sum_{s' \in S} P(s, s') = 1, \forall s \in S_v$. For a given structure, the element $\boldsymbol{P}(s_i, s_j)$ represents the probability that the next state of the process will be $s_j$ given that the current state is $s_i$.
- $L: (S_I \cup S_v) \to 2^{AP}$ is a labeling function. $AP$ is a set of atomic propositions. The labeling function associates to each state the set of atomic propositions that are true in that state.

A state $s \in S$ is said to be an *absorbing state* if $\boldsymbol{P}(s, s) = 1$. If a DTMC contains at least one absorbing state, the DTMC itself is said to be an *absorbing DTMC*. In the simplest model for reliability analysis, the DTMC will have two absorbing states, representing the correct accomplishment of the task and the task's failure, respectively. The use of absorbing states is commonly extended to modeling different failure conditions. For example, different failure states may be associated with different system tasks.

Recalling the behavior models in the above sections, TASD describes the reliability and performance information of system behaviors, the generated LTS captures the processes of system behavior. Hence, we elaborate how to derive the corresponding DTMC by integrating TASD and LTS. The generating algorithm is as follows.

| Algorithm   Generate Variable DTMC |
|---|
| Input : TASD, SAGM |
| Output : DTMC |
| 1:  for all $w_i \in S_{LTS}$ find next state $w_{i+1}$ |
| 2:    if ($|w_{i+1}|=1$) // no branched transitions |
| 3:      if ($|\Sigma(w_i, w_{i+1})| = 1$ & $\Sigma(w_i, w_{i+1})\ not\ for\ selfHealing$) |
| 4:        add $s_i$ to $S_I$, label $\Sigma(w_i, w_{i+1})$ to $s_i$, add $f_{si}$ to $S_I$ // $f_{si}$ is the failure state related to $s_i$ |
| 5:      end if |
| 6:      if ($|\Sigma(w_i, w_{i+1})| > 1$ & $\Sigma(w_i, w_{i+1})\ not\ for\ selfHealing$) |
| 7:        add $s_i$ to $S_v$, label $\Sigma(w_i, w_{i+1})$ to $s_i$, add $f_{si}$ to $S_I$ |
| 8:      end if |
| 9:      if ($\Sigma(w_i, w_{i+1})\ for\ selfHealing$) |
| 10:       add $h_i$ to $S_I$, label $\Sigma(w_i, w_{i+1})$ to $h_i$, add $f_{hi}$ to $S_I$ // $f_{hi}$ is the failure state related to $h_i$ |
| 11:     end if |
| 12:     if ($s_i$ is initial state) |
| 13:       $s_0=s_i$, find o performing $L(s_0)$ in TASD, add $T(s_0,f_{s0})$ with $P(s_0,f_{s0}) = $TAG(o, FP) |
| 14:     else if ($s_i$ is not the last state) |
|           find $o_i$ performing $L(s_i)$ in TASD, add $T(s_i,f_{si})$ with $P(s_i,f_{si}) = $ TAG($o_i$, FP) |
|           add $T(s_i, s_{i-1})$ with $P(s_i, s_{i-1}) = 1-$TAG($o_{i-1}$, FP) // $o_{i-1}$ performs $L(s_{i-1})$ |
| 15:       //if $s_i \in S_v$  $P(s_i,f_{si})$ and $P(s_i, s_{i-1})$ are vectors of probabilities |
| 16:     else |
| 17:       add self-loop with $P(s_i, s_i)=1$ //$s_i$ is an absorbing state |
| 18:     end if |
| 19:     if ($h_i$ is initial state of selfHealing process) |
|           substitute the related failure state with $h_0$ and keep the incoming transition |
| 20:       find o performing $L(h_0)$ in TASD, add $T(h_0,f_{h0})$ with $P(h_0,f_{h0}) = $ TAG(o, FP) |
| 21:     else if ($h_i$ is not the last state of selfHealing process) |
|           find $o_i$ performing $L(h_i)$ in TASD, add $T(h_i,f_{hi})$ with $P(h_i,f_{hi})$ of TAG($o_i$, FP) |
| 22:       add $T(h_i, h_{i-1})$ with $P(h_i, h_{i-1}) =1-$TAG($o_{i-1}$, FP) // $o_{i-1}$ performs $L(h_{i-1})$ |
| 23:     else |
|           find $o_i$ performing $L(h_i)$ in TASD, add $T(h_i,f_{hi})$ with $P(h_i,f_{hi}) = $ TAG($o_i$, FP) |
| 24:       add $T(h_i, h_{i-1})$ with $P(h_i, h_{i-1}) = 1-$TAG($o_{i-1}$, FP) |
| 25:       add $T(h_i, s_{heal})$ with $P(h_i, s_{heal}) = 1-$TAG($o_i$, FP) |
| 26:     end if |
| 27:   else |
| 28:     for each branches perform Algorithm 1 |
| 29:   end if |

| | |
|---|---|
| **30:** | end for |
| **31:** | return $S_I, S_V, s_0, P, L$ |

In our MobIS example, after applying the generating algorithm, we can derive the variable DTMC presented in Figure 13. The gray-colored states refer to the variable state. The transition probabilities from these states are probability vectors. This DTMC contains five absorbing states, including four failure states and a success state. Though, state 15 has two outgoing transitions, system should choose one path for operation at runtime. Hence, the adaptation decision should be made by configuring each variable state.

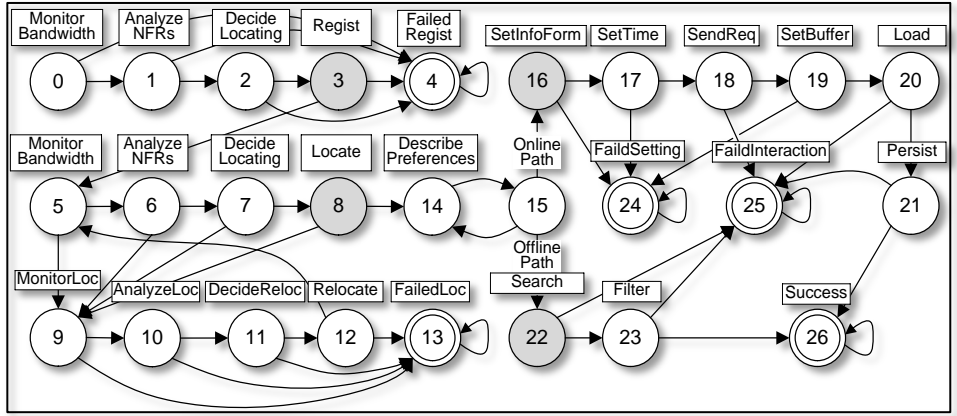

**Figure 13 Variable DTMC of the MobIS**

The failure probabilities of transitions are provided in Table 3 and Table 4. Specifically, if the failure probability of $S_i$ is FP, the success probability is 1-FP. For example, $FP_{s0}$ =0.0005 means that $P(s_0,s_4)$=0.0005 while $P(s_0,s_1)$=0.9995. The failure probabilities of variable states are identified according to context, as discussed in Section 5.1.

**Table 3 Failure probabilities (FP) of invariable states**

| S | FP | S | FP | S | FP | S | FP | S | FP | S | FP |
|---|---|---|---|---|---|---|---|---|---|---|---|
| $S_0$ | 0.0005 | $S_1$ | 0.0001 | $S_2$ | 0.0002 | $S_5$ | 0.0005 | $S_6$ | 0.0001 | $S_7$ | 0.0004 |
| $S_9$ | 0.0001 | $S_{10}$ | 0.0015 | $S_{11}$ | 0.0005 | $S_{12}$ | 0.001 | $S_{14}$ | 0 | $S_{15}$ | 0.05 |
| $S_{17}$ | 0.0005 | $S_{18}$ | 0.005 | $S_{19}$ | 0.0005 | $S_{20}$ | 0.005 | $S_{21}$ | 0.0002 | $S_{23}$ | 0.0005 |

**Table 4 Failure probabilities (FP) of variable states**

| Context = C1 (Bandwidth <=100kbps) | | | | | | | | |
|---|---|---|---|---|---|---|---|---|
| S | $S_3$ | | $S_8$ | | $S_{16}$ | | $S_{22}$ | |
| Option | SMS | Internet | GPS | GSM | Media | Text | Voice | Text |
| FP | 0.02 | 0.1 | 0.05 | 0.15 | 0.08 | 0.02 | 0.05 | 0.02 |
| Context = C2 (Bandwidth >100kbps) | | | | | | | | |
| S | $S_3$ | | $S_8$ | | $S_{16}$ | | $S_{22}$ | |
| Option | SMS | Internet | GPS | GSM | Media | Text | Voice | Text |
| FP | 0.15 | 0.05 | 0.1 | 0.02 | 0.05 | 0.01 | 0.05 | 0.02 |

## 7.2 Reliability Requirements

Formal languages to express properties of systems modeled through DTMCs have been studied in the past and several proposals are supported by model checkers to prove that a model satisfies a given property. In this paper, we focus on Probabilistic Computation Tree Logic (PCTL) [32], a logic that can be used to express a number of reliability properties.
PCTL is defined by the following syntax:

$$\Phi ::= \text{true} \mid a \mid \Phi \land \Phi \mid \neg \Phi \mid \mathcal{P}_{\bowtie p}(\Psi)$$
$$\Psi ::= X\Phi \mid \Phi U^{\leq t}\Phi$$

where $p \in [0,1]$, $\bowtie \in \{<, \leq, >, \geq\}$, $t \in \mathbb{R}_{\geq 0}$ and *a* represents an atomic proposition. The temporal operator *X* is called *Next* and *U* is called *Until*. Formulae generated from $\Phi$ are referred to as *state formulae* and they can be evaluated to either true or false in every single state, while formulae generated from $\Psi$ are named path formulae and their truth is to be evaluated for each execution path. Semantic details of state formulae and path formulae are omitted here for simplicity.
From the *Next* and *Until* operators it is possible to derive others. For example, the *Eventually* operator (often represented by the $\diamond^{\leq t}$ symbol) is defined as: $\diamond^{\leq t}\Phi \equiv true\ U^{\leq t}\Phi$. It is customary to abbreviate $U^{\leq \infty}$ and $\diamond^{\leq \infty}$ as $U$ and $\diamond$.

PCTL can naturally represent reliability-related properties for a DTMC model of the application. For example, we may easily express constraints that must be satisfied concerning the probability of reaching absorbing failure or success states from a given initial state. These properties belong to the general class of reachability properties. Reachability properties are expressed as $\mathcal{P}_{\bowtie p}(\diamond \Phi)$, which expresses the fact that the probability of reaching any state satisfying $\Phi$ has to be in the interval defined by constraint $\bowtie p$. In most cases, $\Phi$ just corresponds to the atomic proposition that is true only in an absorbing state of the DTMC. In the case of a failure state, the probability bound is expressed as $\leq x$, where $x$ represents the upper bound for the failure probability; for a success state it would be instead expressed as $\geq x$, where $x$ is the lower bound for success.

To derive the adaptation decisions in our approach, we consider two reliability requirements in the MobIS example, which can be described with PCTL:
- **R1**: $P_{\geq 0.85}[\ true\ U\ "success"]$ it refers to "*Probability of success shall be greater than 0.85*".
- **R2**: $P_{\geq 0.90}[\ "locate"\ U\ "success"]$ it refers to "*Probability of success after locate task shall be greater than 0.9*".

## 8. Performance- related Adaptation

This section provides the second phase of adaptation, which is performed based on performance requirements. Similar to Section 7, we first describe the leveraged behavior model and then the concerned requirements.

### 8.1 Real-time Behavior Model

Continuous Time Markov Chains (CTMCs) are stochastic models that allow one to express the execution time of each step of the process. Each execution time is characterized by means of an exponential distribution. The classical formalization of CTMCs provided in [33]. However, except for the execution time, we consider the performance about utility and energy costs. Meanwhile, alternative structures and behaviors should also be aggregated into the CTMC. Hence, we extend a CTMC with *variable states* and *rewards*. A Rewarded Variable CTMC is defined as a tuple $(S_I, S_v, s_0, R, \mathcal{R}, L)$ where:
- $S_I$, $S_v$, $s_0$ and $L$ are defined as for DTMCs.
- $R: (S_I \cup S_v) \times S_I \to \mathbb{R}_{\geq 0}$ is the set of *rate* matrix. For a given structure, the element $R(s_i, s_j)$ represents the rate at which the process moves from state $s_i$ to state $s_j$.
- $\mathcal{R}: (S_I \cup S_v) \times Tr \to \mathbb{R}_{\geq 0}$ is the set of tagged rewards that assign a non-negative real value to transition $Tr$.

An *absorbing state* $s_i$ is a state that $R(s_i, s_j)=0$ for all states $s_j$. For any pair of states $(s_i, s_j)$, $R(s_i, s_j)>0$ iff there exists a transition from $s_i$ to $s_j$. Furthermore, $1 - e^{-R(s_i, s_j) \cdot t}$ is the probability that the transition from state $s_i$ to state $s_j$ is taken within $t$ time units. Through rates, we model the delay in choosing the transition. If more than one transition exiting state $s_i$ has an associated positive rate, a *race* among transitions occurs. The probability $P(s_i, s_j)$ that control, held by $s_i$, will be transferred to state $s_j$ corresponds to the probability that the delay of going from $s_i$ to $s_j$ is smaller than the one of going toward any other state; formally, $P(s_i, s_j) = R(s_i, s_j)/E(s_i)$, where $E(s_i) = \sum_{s_j \in S} P(s_i, s_j)$. $E(s_i)$ is called exit rate of state $s_i$.

Based on the derived DTMC, it is easy to generate the corresponding CTMC with the following steps:
• Step 1: Remove all the absorbing states that depict task failures, i.e. only the *success* state remains.
• Step 2: Find related *time costs* (CT) in tags of TASD and assign 1/TC to related transitions as transition rate.
• Step 3: Find related *utility* (U) and *energy costs* (EC) in tags of TASD and assign them to related state as rewards.

The generated CTMC is presented in Figure 14. The gray-colored states also represent alternative structures. The transition rates from these states are also vectors. Similarly, the adaptation decision should also be made by configuring each variable state. The concrete values of transition rate, energy costs and utility are presented in Table 5 and Table 6. Notice that some transitions have no assigned values of utility, because these transitions refer to failures of tasks, e.g. $T_{4,8}$ refers to the failure of *monitorBandwidth*, the utility is assigned to 0.

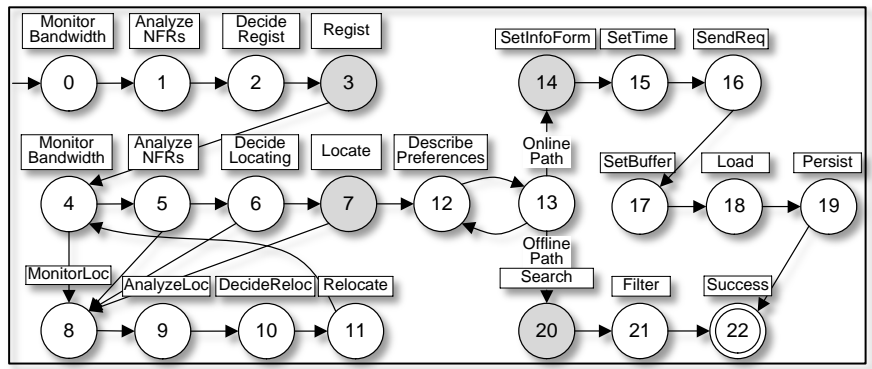

**Figure 14 Variable CTMC of the MobIS**

**Table 5 Utility(U), Energy Costs (EC) and Rate (R) of transitions from invariable states**

| T | U | EC | R | T | U | EC | R | T | U | EC | R |
|---|---|---|---|---|---|---|---|---|---|---|---|
| $T_{0,1}$ | 0.1 | 3 | 1/0.5 | $T_{1,2}$ | 0.2 | 5 | 1/0.8 | $T_{2,3}$ | 0.1 | 5 | 1/1 |
| $T_{4,5}$ | 0.15 | 3 | 1/0.5 | $T_{5,6}$ | 0.2 | 5 | 1/0.8 | $T_{6,7}$ | 0.1 | 5.5 | 1/1.2 |
| $T_{4,8}$ | 0 | 3 | 1/3 | $T_{5,8}$ | 0 | 5 | 1/5 | $T_{6,8}$ | 0 | 5.5 | 1/8 |
| $T_{8,9}$ | 0.1 | 1 | 1/0.1 | $T_{9,10}$ | 0.1 | 6 | 1/1 | $T_{10,11}$ | 0.5 | 2 | 1/0.5 |
| $T_{11,4}$ | 2 | 3 | 1/0.2 | $T_{12,13}$ | 0.2 | 5 | 1/15 | $T_{13,12}$ | 1 | 2 | 1/10 |
| $T_{13,14}$ | 0.1 | 2 | 1/1.5 | $T_{13,20}$ | 0.3 | 2 | 1/1.5 | $T_{15,16}$ | 0.1 | 3 | 1/0.5 |
| $T_{16,17}$ | 0.5 | 20/T | 1/(2/T) | $T_{17,18}$ | 0.1 | 3 | 1/0.5 | $T_{18,19}$ | 0.2 | 20*B | 1/(20/B) |
| $T_{19,22}$ | 0.1 | 5 | 1/0.8 | $T_{21,22}$ | 0.2 | 4 | 1/0.5 | | | | |

**Table 6 Utility(U), Energy Costs (EC) and Rate (R) of transitions from variable states**

| Context = C1(Bandwidth <=100kbps) | | | | | | | | | | | | |
|---|---|---|---|---|---|---|---|---|---|---|---|---|
| T | $T_{3,4}$ | | $T_{7,8}$ | | $T_{7,12}$ | | $T_{14,15}$ | | $T_{20,22}$ | | $T_{16,17}$ | $T_{18,19}$ |
| Option | SMS | Internet | GPS | GSM | GPS | GSM | Media | Text | Voice | Text | T | B |
| U | 6 | 10 | 0 | 0 | 15 | 6 | 15 | 12 | 15 | 5 | | |
| EC | 3 | 6 | 8 | 7 | 8 | 7 | 12 | 6 | 5 | 2 | 20/T | 20*B |
| R | 1/10 | 1/5 | 1/10 | 1/20 | 1/5 | 1/10 | 1/10 | 1/6 | 1/1 | 1/0.5 | 1/(2/T) | 1/(20/B) |
| Context = C2(Bandwidth >100kbps) | | | | | | | | | | | | |
| T | $T_{3,4}$ | | $T_{7,8}$ | | $T_{7,12}$ | | $T_{14,15}$ | | $T_{20,22}$ | | $T_{16,17}$ | $T_{18,19}$ |
| Option | SMS | Internet | GPS | GSM | GPS | T | B | Text | Voice | Text | T | B |
| U | 5 | 8 | 0 | 0 | 10 | | 20 | 9 | 15 | 5 | | |
| EC | 3 | 6 | 15 | 3 | 15 | 20/T | 8 | 4 | 5 | 2 | 20/T | 20*B |
| R | 1/4 | 1/2 | 1/15 | 1/15 | 1/8 | 1/(2/T) | 1/6 | 1/3 | 1/1 | 1/0.5 | 1/(2/T) | 1/(20/B) |

## 8.2 Performance Requirements

In this paper, we consider Continuous Stochastic Logic [33] to state properties on CTMC models. For CTMC there are two main type of properties relevant for analysis: *steady-state* properties where the system is considered "on the long run", that is, when an equilibrium has been reached; *transient* properties where the system is considered at specific time points or intervals. CSL is able to express both steady-state and transient properties by means of the $\mathcal{S}$ and $\mathcal{P}$ operators, respectively. Besides, to compute the reward, we extend the state formulae with related syntax. The syntax of extended CSL is recursively defined as follows:

$\Phi ::= \text{true} \mid a \mid \Phi \wedge \Phi \mid \neg \Phi \mid \mathcal{S}_{\bowtie p}(\Phi) \mid \mathcal{P}_{\bowtie p}(\Psi) \mid \mathcal{R}_{\bowtie r}(\Psi) \mid \mathcal{R}_{\bowtie r}(I^{=t}) \mid \mathcal{R}_{\bowtie r}(C^{\leq t})$

$\Psi ::= X^{\leq t}\Phi \mid \Phi U^{\leq t}\Phi$

where $p \in [0,1]$, $\bowtie \in \{<, \leq, >, \geq\}$, $t \in \mathbb{R}_{\geq 0} \cup \{\infty\}$ and *a* represents an atomic proposition.

$\mathcal{S}_{\bowtie p}(\Phi)$ describes the probability of being in a state where $\Phi$ holds when $t \to \infty$, evaluated on all paths originating in *s*, meets the bound $\bowtie p$. $\mathcal{P}_{\bowtie p}(\Psi)$ refers to the probability that the set of paths starting in *s* and satisfying $\Psi$ meets the bound $\bowtie p$. $\mathcal{R}_{\bowtie r}(\Psi)$ describes the reward that the set of paths starting in *s* and satisfying $\Psi$ meets the bound $\bowtie r$. $\mathcal{R}_{\bowtie r}(I^{=t})$ depicts the instantaneous reward at time t meets the bound $\bowtie r$. $\mathcal{R}_{\bowtie r}(C^{\leq t})$ means the cumulative reward until time t meets the bound $\bowtie r$. With the above syntax, three considered performance requirements of the MobIS example can be specified as follows:

- **R3**: $R\{"utility"\}_{\geq 30}[\text{ true } U \text{ "success"}]$ it refers to "*Utility of success shall be greater than 30*".
- **R4**: $P_{\geq 0.85}["locate" U^{\leq 50} "success"]$ it refers to "*Probability of success from locate task to wait less than 50s is greater than 0.85*".
- **R5**: $R\{"energy"\}_{\leq 180}[C^{\leq 50}]$ it refers to "*Accumulative energy costs in 50s shall be less than 180*".

## 9. Evaluation Results

In the above two sections, we have derived the behavior models for computing reliability requirements and performance requirements. This section presents the evaluation of achieving adaptation against these non-functional requirements via verification. Specifically, the adaptation decision should be made for satisfying each of the five NFRs. To this end, we conduct the verification process for each possible path in the DTMC and CTMC. After the verification of each NFR, the number of candidate solutions will reduce accordingly.

For implementing verification, we consider taking the advantage of probabilistic model checking [34], which is an automatic technique for verifying quantitative properties for probabilistic systems. To this end, our approach adopts a well-established stochastic model checker PRISM [35, 36] to support the model checking process.

We performed the verification with PRISM 4.3 on a machine with an AMD FX8350 (3.8GHz) CUP with 4GB of RAM. The quantitative settings of DTMC and CTMC have been provided in Table 3 to Table 6. The values of time interval ($T$) and buffer ($B$) are chosen from 1to 20 with step size of 2. Hence, the MobIS totally includes 808 alternative adaptation decisions for each kind of contexts. The results presented below depict the reduction of decision space after verifying each NFR in context C2 ((Bandwidth>100kbps)).

Figure 15 depicts six adaptation candidates are chosen by verifying reliability requirement *R1*. The optional structures can be found in the legend. The probabilities in legend order are 0.8670, 0.8899, 0.9268, 0.9000, 0.9238 and 0.9529.

Figure 16 presents four of the six candidates are selected as adaptation decision by verifying reliability requirement *R2*. The probabilities in legend order are 0.9196, 0.9583, 0.9296 and 0.9687.

By verifying *R3*, only the structure {Internet, GSM, Online (Media)} is chosen as the adaptation structure, as presented in Figure 17. It is interpreted as when bandwidth is greater than 100kbps, the user shall register through Internet, be located with GSM and receive pushed media information from business server.

By verifying *R4*, candidates of parameters are determined. In Figure 18, 21 pairs of ($T$, $B$) are selected: (5,19), (7,17), (7,19), (9,15), (9,17), (9,19), (11,15), (11,17), (11,19), (13,15), (13,17), (13,19), (15,15), (15,17), (15,19), (17,15), (17,17), (17,19), (19,15), (19,17) and (19,19).

After verifying *R5*, as depicted in Figure 19, 10 pairs of ($T$, $B$) satisfy the energy requirement, including (5,19), (7,17), (9,15), (9,17), (11,15), (11,17), (13,19), (15,19), (17,19) and (19,19). Among them, we choose the parameters that cost least energy. Then the time interval is assigned to 9 and buffer is set to 15.

Thus, after verifying each alternative path of the DTMC and CTMC against the NFRs, the optimal adaptation decision is derived as {RegistOption=byInternet, LocateOption=byGSM, InfoOption=Online, InfoFormOption=Media, TimeInterval=9, Buffer=15}.

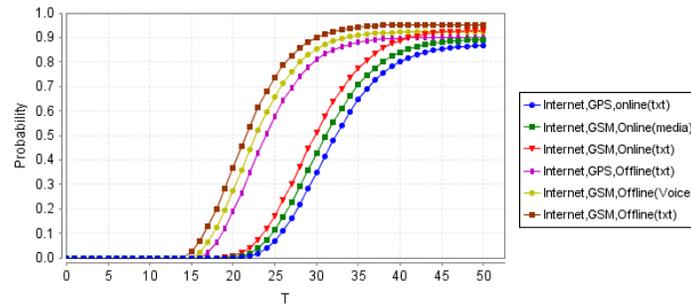

**Figure 15 Adaptation decision chosen by verifying R1**

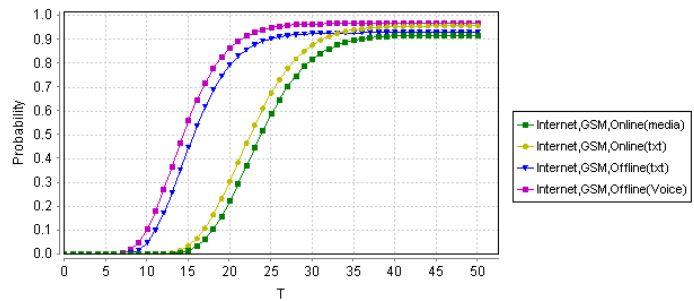

**Figure 16 Adaptation decision chosen by verifying R2**

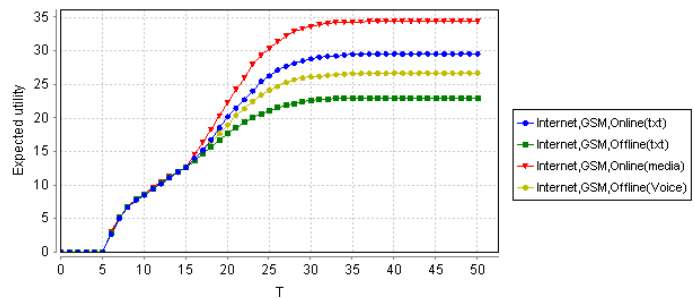

**Figure 17 Adaptation decision chosen by verifying R3**

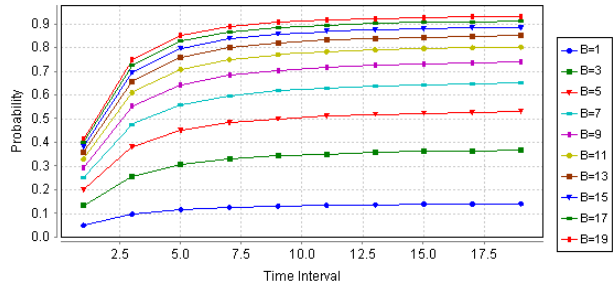

**Figure 18 Adaptation decision chosen by verifying R4**

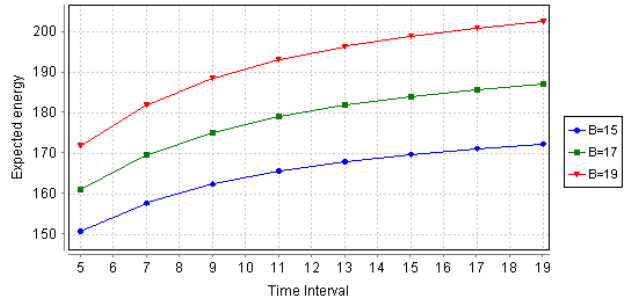

**Figure 19 Adaptation decision chosen by verifying R5**

## 10. Related Work

**Modeling adaptation mechanism.** RE for self-adaptive systems demands modeling both business logic and adaptation logic. Qureshi et al. [6] focuses on the question about how to represent requirements for SAS in a way which can be read by non-engineering stakeholders. They propose a modeling language, called Adaptive RML, for the representation of early requirements for SASs. The language has graphical primitives in line with classical goal modeling languages. Cheng et al. [7] introduces a goal-based modeling approach to develop the requirements for a DAS, while explicitly factoring uncertainty into the process and resulting requirements. They introduce a variation of threat modeling to identify sources of uncertainty and demonstrate how the RELAX [10] specification language can be used to specify more flexible requirements within a goal model to handle the uncertainty. As the following work, Baresi et al. [37] propose an approach to represent requirements for service compositions by extends traditional goal models with adaptive goals to support continuous adaptation. They also provide a methodology to trace goals onto the underlying composition, assess goals satisfaction at runtime, and activate adaptation consequently. In comparison with their work, our approach describes each adaptation mechanism as a MAPE loop the modeling process can be manipulated conveniently. The syntax of the modeling method is brief but efficient.

**Achieving adaptation.** To achieve adaptation, researchers proposed both computing-based methods and reasoning-based method. FUSION was proposed by Elkhodary *et al*. [11]. The approach uses online learning to mitigate the uncertainty associated with changes in context and tune system behaviors to unanticipated changes. Esfahani *et al*. [12] proposed POISED for improving the quality attributes and achieve a global optimal configuration of a system by assessing both the positive and negative consequences of context uncertainty. Wang *et al*. [13] focused on monitoring and analysis aspect. They proposed a framework for diagnosing failure of software requirements by transforming the diagnostic problem into a propositional satisfiability problem, which is solved by SAT solvers. In [14], Wang and Mylopoulos proposed an autonomic architecture consisting of monitoring, diagnosing, reconfiguration and execution component. Comparing with their work, our approach provides a different way to deriving adaptation decision, i.e. by verification. This process is completed though Bayesian reasoning with structured behavior models, i.e. DTMCs and CTMCs. In our approach, verification is used as a decision-making process. Besides, we take into consideration several kinds of NFRs.

**Requirements verification.** Goldsby et al. [15] provide AMOEBA-RT, a run-time monitoring and verification technique that provides assurance that dynamically adaptive software satisfies its requirements. Filieri and Tamburrelli [16] provide a model checking based approach for verification, in which reliability models are given in terms of Discrete Time Markov Chains which are verified against a set of requirements expressed as logical formulae. Epifani et al. [17] lay the foundations for an iterative model-driven development, which aims at verifying that an implementation satisfies non-functional requirements. In their approach, if the resulting running system behaves differently from the assumptions made at design time, the feedback to the model shows why it does not satisfy the requirements and lead to a further development iteration. Ghezzi et al. [18] put forward the proposal of quantitative verification at runtime for self-adaptive service-based software. Ghezzi and Sharifloo [19] propose a model-based approach that enables software engineers to assess their design solutions for software product lines in the early stages of development. Ghezzi and Tamburrelli [38] present a method and supporting tools to rea-son about requirements of open-world systems built as integrated services that rely on existing, externally provided services. Besides, Zhao et al. [39] propose the mode-supported Linear Temporal Logic that is an effective way to describe global specifications of adaptive software. The global specifications are defined for adaptive software as requirements from the perspective of global adapting process. The model checking problem is also resolved using Linear Temporal Logic and Labelled Transition System Analyzer. In our approach, we also focus on verification of NFRs. However, we concern more system properties, i.e. utility, time costs and energy costs. Meanwhile, we apply verification as the decision-making method during self-adaptation.

## 11. Conclusions and Future Work

In this paper, we proposed a model-driven approach to assuring non-functional requirements through runtime adaptation, which is implemented via a series of verification processes. We use the adaptation goal model to describe both business logic and adaptation logic. By tagging reliability and performance to system tasks, we define the related tagged adaptation goal model. The behaviors of adaptation tasks are specified in the tagged adaptation sequential diagrams. To formalize system behaviors, we propose a method for transforming the requirements model to behavior model that is expressed as a LTS. The sequential diagram and LTS are merged into a variable DTMC and a rewarded variable CTMC. Related NFRs are specified as PCTL and CSL properties. The verification is implemented within PRISM, a tool for probabilistic model checking. The approach has been successfully applied to a mobile information system for achieving adaptation decision.

The approach can serve as a general guidance for achieving adaptation of SASs in several aspects:

1. The adaptation goal model provides a graphical representation about adaptation mechanism, which can be easily manipulated by adding MAPE tasks to original goal model. Though the representation is concise, it is powerful for generating the behavior models related to these mechanisms.
2. The tagged adaptation goal model is generated by reliability and performances attributes. Notice that the key idea implied here is these attributes are varying according to the operating environment and system configurations. This occurs in many software-intensive systems and it enables the alternative adaptation decisions.
3. The process of transforming a goal model to its behavior model can be applied to SASs of any other domains. It provides pattern-based CSP-specifications, which support simulation and verification within analysis toolkits. Meanwhile, the verification results reveal the correctness of behavior models.
4. Though requirements verification is discussed a lot by the RE community. Few works focus on the performance of utility, time costs and energy costs. Our approach proposes a solution to these research gaps. Another key point should be noticed is that we derive adaptation decision by selecting from the verification results. In other words, the verification processes serve as the decision-making unit for an optimization problem. That is also the reason why we design the variable states in DTMCs and CTMCs. Actually, different paths of DTMCs and CTMCs characterized by the variable states are treated as alternative decision-paths.
5. The whole approach starts from RE stage and is highly based on models, including static models and dynamic (behavior) models. It may help researchers and practitioners who focus on requirements engineering and model-driven development. In our proposal, by transformation and integration, models in RE stage can help runtime analysis of SASs, i.e. verifying NFRs and achieving adaptation.

Future work includes two directions. First, we concern the other type of uncertainties, i.e. nondeterminism. It characterizes a system that has two or more behaviors at a certain time point. For example, when a request comes into the server queue, the server will either serve the request or wait for another request. This situation is different from the probabilistic attributes of system behaviors. The nondeterminism can be modeled in Markov Decision Processes (MDPs) appropriately. We are interested in how to achieve adaptation decision for the nondeterministic adaptive systems, especially for assuring NFRs. We also concern the verification of NFRs with other uncertain attributes, especially the flexibility provided by introducing fuzziness into NFRs. To this end, we need to build effective adaptation loops, e.g. MAPE loops and reasonably describe contextual changes, fuzzy requirements and adaptation logic. We consider UPPAAL-SMC as a proper solution to this problem.